Reconciling Trends in U.S. Male Earnings Volatility:

Results from Survey and Administrative Data

Robert Moffitt
Johns Hopkins University

John Abowd
U.S. Bureau of the Census
and Cornell University

Christopher Bollinger
University of Kentucky

Michael Carr
University of Massachusetts Boston

Charles Hokayem
U.S. Bureau of the Census

Kevin McKinney
U.S. Bureau of the Census

Emily Wiemers
Syracuse University

Sisi Zhang
Jinan University

James Ziliak
University of Kentucky

January 31, 2022

The authors would like to thank Joe Altonji, the participants in a Summer 2019 meeting sponsored by the Washington Center for Equitable Growth, the participants of the Summer 2020 NBER Conference in Research and Wealth workshop, and the participants of the June 2021 Summer meetings of the Econometric Society for comments. This draft also incorporates suggestions made by the JBES editorial staff and referees on individual papers.

Abstract

Reconciling Trends in U.S. Male Earnings Volatility:
Results from Survey and Administrative Data

There is a large literature on earnings and income volatility in labor economics, household finance, and macroeconomics. One strand of that literature has studied whether individual earnings volatility has risen or fallen in the U.S. over the last several decades. There are strong disagreements in the empirical literature on this important question, with some studies showing upward trends, some showing downward trends, and some showing no trends. Some studies have suggested that the differences are the result of using flawed survey data instead of more accurate administrative data. This paper summarizes the results of a project attempting to reconcile these findings with four different data sets and six different data series--three survey and three administrative data series, including two which match survey respondent data to their administrative data. Using common specifications, measures of volatility, and other treatments of the data, four of the six data series show a lack of any significant long-term trend in male earnings volatility over the last 20-to-30+ years when differences across the data sets are properly accounted for. A fifth data series (the PSID) shows a positive net trend but small in magnitude. A sixth, administrative, data set, available only since 1998, shows no net trend 1998-2011 and only a small decline thereafter. Many of the remaining differences across data series can be explained by differences in their cross-sectional distribution of earnings, particularly differences in the size of the lower tail. We conclude that the data sets we have analyzed, which include many of the most important available, show little evidence of any significant trend in male earnings volatility since the mid-1980s.

The literature on labor market volatility is vast and touches on multiple areas of macroeconomics, household finance, labor economics, and overlaps between them. The classic study of permanent vs transitory components of income and their implications for consumption, saving, and the marginal propensity to consume is just one example (Friedman, 1957; Hall and Mishkin, 1982). On the micro level, this literature has spilled over into household finance, with its concern with liquidity constraints, ability to deal with income shocks, possible inadequacy of assets to deal with such shocks, and consequent inability to smooth consumption sufficiently (Carroll, 1997; Gourinchas and Parker, 2002; Blundell et al., 2008; Ganong and Noel, 2019). In labor economics, a literature going back to the 1960s and 1970s on sectoral models of the labor market, with one sector characterized by high wages and stable jobs and another characterized by low wages and unstable jobs, has reemerged in recent discussions of technological change and the decline of union and manufacturing jobs, since the latter are generally more stable than average (Taubman and Wachter, 1986; Katz and Autor, 1999). The impact of income uncertainty on investments in human capital, both educational and on-the-job, and on investments in children at young ages, has generated yet another discussion in labor economics (Levhari and Weiss, 1974; Cunha et al., 2005; Carneiro and Ginja, 2016).

An important empirical branch of this literature concerns whether volatility has changed over time in the U.S. A priori, whether labor market volatility should be expected to have risen or fallen differs by perspective. On the one hand, the just-mentioned literature on structural change in the U.S. labor market suggests that earnings instability might have increased, at least for workers with medium or lower skills. Katz and Autor (1999), for example, in their review of the early literature on increasing earnings inequality, make the connection between rising earnings inequality and rising instability directly. Haider (2001) also explicitly draws a connection between growing earnings inequality and earnings instability. On the other hand, a prominent hypothesis in macroeconomics is that the 1980s ushered in a period known as the Great Moderation, reflected in declining levels of aggregate volatility (McConnell and Perez-Quiros, 2000). While there is no necessary connection between aggregate volatility and volatility at the micro level (as noted by Davis and Kahn, 2008, and Dynan et al., 2012), some macroeconomists argue that a lack of decline in individual earnings volatility matching the aggregate volatility decline is intuitively difficult to explain (Sabelhaus and Song, 2010).

The project which this Overview summarizes represents an effort to bring several data sets to bear on the question of whether U.S. earnings volatility at the micro level has risen, fallen, or remained constant over the last several decades. It is motivated in large part by the disparate findings on this question which have appeared when different data sets have been used. The workhorse data set for estimating trends in individual earnings volatility in the U.S. has been the Panel Study of Income Dynamics (PSID), a longitudinal survey that has been ongoing since 1968 (and is hence the longest-running general-purpose socioeconomic panel in the world), which has attempted to maintain reasonable population representativeness and which asks extensive questions on labor market activity. The use of the PSID for the study of male earnings volatility began with Gottschalk and Moffitt (1994), who found male earnings volatility to have increased from 1970 to 1987, with the largest increase occurring among the less educated. About a dozen PSID studies subsequent to the Gottschalk-Moffitt study have also found increases in male earnings volatility over time.[1] However, as we will discuss in detail in the first section of the paper below, findings have often differed in other data sets. While some differing findings have been found in other survey data sets (e.g., the Survey of Income and Program Participation and, partially, the Current Population Survey), the largest differences have emerged from studies using administrative data from Social Security, tax, or Unemployment Insurance records, which often find no trend in earnings volatility or even a decline. We review these studies in detail in the first section of the paper below. The difference in trends found in administrative data, which are often presumed to be more accurate than survey data, suggests that the PSID may be biased by reporting error, attrition bias, or some other issue.

---

[1] See Dynarski and Gruber (1997), Haider (2001), Hyslop (2011), Keys (2008), Heathcote et al. (2010), Shin and Solon (2011), Dynan et al. (2012), Moffitt and Gottschalk (2012), Jensen and Shore (2015), and Carr and Wiemers

The project brings four different subprojects and six different data series to bear on this question, each using common samples to the maximum extent feasible, common definitions of volatility, and common other treatments of the data (trimming of outliers, treatment of nonworkers, imputations, and others). One paper reexamines the oft-used PSID, but adds to previous work by extending the time period up through 2016--which turns out to be important-- and by conducting a number of analyses of bias that might come from attrition and other threats to representativeness. A second paper uses data from the Current Population Survey, using earnings reports of individuals who are in the survey twice over a two-year period. But this paper also links the CPS sample to Social Security earnings files, permitting a direct comparison of survey reports and administrative data reports for the same individuals. The third paper uses the Survey of Income and Program Participation (SIPP), a well-known Census Bureau survey intended to be representative of the population and which consists of a rolling series of 2-to-5-year panels, thereby permitting estimates of the volatility of year-to-year changes in earnings. But, like the second paper, this third paper also employs a data set of Social Security earnings

---

(2018). A full listing of all studies, along with those using other data sets, can be found in Moffitt and Zhang (2018).

data matched to the SIPP survey respondents, again providing the opportunity to compare trends in earnings volatility between the two types of data. The fourth paper uses only administrative data drawn from Unemployment Insurance (UI) wage records collected by employers and reported to state governments. The well-known file is called the Longitudinal Employer-Household Dynamics (LEHD) data set and has the near universe of earnings of UI-covered workers in the states that have provided their data. The importance of the LEHD is not only its vast sample size compared to the survey-based data sets, but its different sampling frame—namely, the near-total universe of US workers and not just those who agreed to participate in a survey.

The four papers each focus on trends in male earnings volatility over the years for which they have data. The PSID goes back to 1970, the SIPP goes back to the early 1980s, the matched CPS data we have goes back to 1996, and the LEHD goes back to 1998. There is thus full overlap after 1998 and partial overlap in many earlier years. A summary of the analyses and findings of each paper is given in this Overview, and the four individual papers which follow provide additional detail on each subproject.

The major findings of the analysis are that, when treated on a comparable basis, there is considerable agreement in volatility trends across the data sets, although less in levels. In terms of levels, the LEHD has the highest volatility and, for the two data sources where linked administrative and survey data are available, the administrative data show higher levels of volatility than the survey data. In terms of trends, we summarize results for three successive time periods separately. First, while none of the data sets reach back as far in time as the PSID, two that reach back to the 1970s or early 1980s (the SIPP survey and administrative data) are consistent with the PSID by showing some upward trends in volatility over that period. But, second, both the PSID and those two data sets also show no average trend from the mid-1980s to the late 1990s. Third, after the late 1990s, all six data series are available and all show countercyclical volatility patterns, rising just before and in the early phases of the Great Recession and falling afterwards. However, with the exception of the PSID and the LEHD, all data sets show no net trend from the late 1990s to the mid- or late 2010s after the Recession was over. The PSID shows a positive net trend, but smaller than that which occurred in the 1970s-1980s, and the LEHD shows a small negative net trend after 2011. We conclude that the evidence shows no strong overall trend in volatility among working men in the U.S. since the mid-1980s, i.e., over the last 30 years.

While we conduct a number of sensitivity tests to the robustness of these findings, by far the most important test we conduct relates to the size of the left tail of the cross-sectional earnings distribution in the different data series. The LEHD has a much larger left tail than the other data sets, and the two data sources that have matched survey and administrative data show larger left tails in the latter than in the former. Other studies have found similar patterns and have often ascribed it to underreporting of short employment spells, which have low earnings, in survey data. We show that when the cross-sectional earnings of the data sets are required to have the same distribution (that of the PSID), the levels of volatility in all the data sets are much closer to each other. Equally important, this exercise converts the small negative trend in the LEHD to a small positive trend and makes the trends in the other administrative data sets over that same period positive instead of zero. The changes are a result of different trends over time in the left tail of the earnings distribution in the PSID and the other data sets   We conclude that the small differences in trends across the data series, particularly in the late periods of our data, are mostly a result of differences in trends in the left tail of earnings.

The next section of the paper reviews the conflicting findings in past work in more detail. The different data sets used in the project appear in the following section, after which the methods and results across the data sets are summarized and attempts at reconciling their differences are reported. A final section summarizes the findings and draws lessons for future work.

I. Past Work: Additional Detail

As noted in the Introduction, there have been over a dozen studies of earnings volatility using the PSID (Moffitt and Zhang, 2018, has a detailed table of those published prior to 2018, describing their samples and results). They do not always align perfectly with each other in methodology, and many estimate error components models instead of the gross volatility models studied here, instead using the transitory variance as the measure of volatility. The studies differ by what years of data were available at that time the studies were conducted. Almost all studies show rising volatility from the 1970s to the 1980s, and either no trend or a downward trend through the 1990s (ignoring cyclical movements). Those studies which had data into the 2000s (e.g., Shin and Solon (2011) and Dynan et al. (2012)) show rising volatility since that time. Carr and Wiemers (2018) had data through 2012 and showed that PSID volatility continued to rise during the Great Recession. Moffitt and Zhang (2018) had data through 2014 and showed that volatility started to decline after the Recession.

The results from other survey data sets are mostly, but not always, roughly consistent with the PSID. Using across-wave matched Current Population Survey (CPS) observations, Celik et al. (2012) also found increases in volatility from the 1970s to the 1980s, stability through the 2000s, then a resumption of an increase, like the PSID. But Ziliak et al. (2011), also using CPS matched data, found the same early trends as Celik et al. but more of a stable trend through the late 2000s, unlike Celik et al. and the PSID. Koo (2016) found similar trends with CPS matched data. Dahl et al. (2008, 2011) examined volatility in household income in the Survey of Income and Program Participation (SIPP) survey and found no trend over time unless imputed income is included. Celik et al. also estimated trends in male earnings volatility in the Survey of Income and Program Participation (SIPP), finding an actual decline after 1984.[2]

Starker differences with the PSID are often found in studies using administrative data. On the one hand, Carr and Wiemers (2018), using Social Security earnings records of SIPP respondents, found patterns similar to the PSID—rising through the early 1980s, declining through 2000, and then rising through the mid-2000s. But Guvenen et al. (2014), using Social Security earnings data not linked to a survey, found slight declines in male volatility from 1980 to 2011.[3] Bloom et al. (2017), also using Social Security earnings records, showed separate volatility trends for men and women and found those for men to decline from 1978 to 2013. Dahl et al. (2008) use the Continuous Work History Sample (CWHS) from Social Security records and find declining male earnings volatility between 1985 and 2003, while Dahl et al. (2011) show declining earnings volatility for all workers in both the CWHS and SIPP-SSA linked data.[4] And using administrative data from Unemployment Insurance records, Celik et al. found no trend in volatility 1992-2008 for the 12 states available in the data, while DeBacker et al. (2013), using tax records, found no trend from 1987 to 2009. Thus the largest discrepancies with the PSID are from those studies using administrative data, with some studies using Social Security data finding declining volatility and those using UI or tax data finding stable rather than increasing volatility.[5]

---

[2] Bania and Leete (2009) examine intrayear volatility with the SIPP for a subset of years.
[3] Sabelhaus and Song (2009,2010), using administrative data from Social Security earnings records, found declining volatility from 1980 to 2005 but on a sample pooling men and women, which is problematic since volatility trends for women appear to be different (more often declining) than those for men, even with the PSID (Dynan et al., 2012).
[4] Their linked sample differed from that of Carr and Wiemers (2018).

There has been little work on reconciling the discrepancies in trends across the data sets. Dahl et al. (2011) suggested that the differences in volatility trends in SIPP survey data and matched Social Security earnings data might be related to imputed earnings values in the SIPP (we carefully examine this issue with the SIPP survey data). Celik et al. presented the diverse findings from different data sets, but had no explanation for the differences.[6] Carr and Wiemers (2018) compared volatility trends in the PSID to those in SIPP-based Social Security earnings data—but not matched to the PSID--and found them to be approximately the same, similar to what we find.

II. The Data Sets

The six data series used in this project are shown in Table 1: the PSID, the CPS survey, CPS linked Social Security earnings records, the SIPP survey, SIPP-linked Social Security earnings records, and UI earnings from the LEHD. The PSID has been analyzed many times before, so the primary purpose of including it is only to provide a baseline estimate using the same sample definitions, measures of volatility, and other analysis features as those in the other five data series.

---

[5] A recent paper (Braxton et al., 2021) uses Social Security earnings records of CPS respondents and finds declining gross volatility from 1982 to 2016, but men and women are combined.
[6] The Celik et al. (2012) study comes closest to this project. The authors find stability of earnings volatility for the CPS and LEHD but a decline for the SIPP and a rise for the PSID in the later periods. They test a number of explanations for the differences with the PSID, finding that none of them adequately explain the difference, similar to our findings. The Celik et al. study only went through 2006 and had a number of other differences in treatment of the data from our study.

For all data sets, only men 25-59 in each year are included.[7]  Regarding sampling frames, five of them are based on survey sampling frames of the non-institutional population and may be subject to nonresponse bias from those declining to participate (Groves et al., 2002; National Research Council, 2013).  In addition, the PSID is only representative of the 1968 US population (at best, ignoring attrition and other issues) because it does not include post-1968 immigrants, as all the other data series do.[8]  The CPS only includes those who were at the same address in the surveys one-year apart (because the Census Bureau just returns to the same address), but the CPS paper finds this restriction to affect only the level of volatility, not its trend (attriters have higher levels of volatility).  The Social Security earnings records matched to the CPS and SIPP surveys are drawn from the same source—the Detailed Earnings Records (DER)—and necessarily exclude those who do not have an SSN, which may include illegal immigrants and some who work off the books.  The LEHD draws its administrative earnings histories from Unemployment Insurance records, and therefore excludes individuals not properly using Social Security numbers as well as those who work off the books, and excludes those who are not covered by the UI system.  These issues are discussed in the LEHD paper.  Most of these differences are unalterable and introduce inevitable noncomparability to some unknown degree.

Regarding the samples, as is well known, the PSID collects sufficient earnings information only on household heads and their spouses. Both the CPS and SIPP surveys have headship information but the SIPP administrative data do not and the LEHD does not because family composition is not collected in UI records, which are reported by employers.[9]  Thus the

---

[7] The CPS paper also reports results for women.
[8] The PSID has made some attempts to bring immigrants back in.  See the PSID paper for a discussion.
[9] The SIPP administrative data set is a separate confidential data set housed at the Census Bureau and, while it represents all those in the survey who were found in the Social Security wage records, is not directly linkable to the

possible importance of headship can be explored in only three of the data series. As for the definition of earnings, the PSID, SIPP survey, and CPS files have wage and salary earnings and self-employment income separately, but the SIPP administrative data and the LEHD contain both and they cannot be separated. Again, some examination of this issue can be conducted but not across all data sets. Estimates of volatility including and excluding nonworkers are conducted in the project but only where possible, for the LEHD data have no nonworkers (in the LEHD, this means the absence of a positive earnings record). The LEHD sample size is vastly larger than that in any of the other data sets, and the PSID sample size is the smallest.[10]

In this Overview paper, we refer to a number of sensitivity tests conducted to gauge the importance of these cross-data differences. However, the details of those tests appear only in the individual papers.

III. Volatility Measures

All analyses use simple and transparent summary measures of gross earnings volatility, calculating the earnings change from one year to a subsequent year, either one or two years later, depending on the data set.[11] The measure we report in this overview paper uses the variance of what is called the arc percent change, which is simply the percent change in earnings relative to the average in the two years.[12] Another common measure is the variance of log earnings differences, but this measure is more sensitive to the tails. But the individual papers report results for that measure as well and find little difference in estimated trends. No attempt is

---

individual survey records, and the Bureau does not put survey-reported headship status on the file.
[10] The LEHD data do not have all states. The LEHD paper provides evidence that the states it has are representative of all states.
[11] In 1997, the PSID went to every-other-year interviewing. All PSID results are calculated over 2-year intervals for all periods to insure comparability.
[12] This measure is commonly used in macroeconomic studies (e.g., Davis et al., 2006).

made to decompose the variance of earnings changes into permanent and transitory components; this is left for future work.[13] Also, while the results shown in this Overview paper are based on volatility estimated with earnings directly, the individual papers also show results for volatility calculated using residuals from a regression of either the change in log earnings or the arc percent change on age and age squared, also common in the literature. No difference in volatility trends is found when using such residuals. The baseline specification also trims the top and bottom one percent of the cross-sectional earnings distribution in each year to remove outliers, but sensitivity tests to trimming are conducted and, in fact, differences in the sizes of the tails of the earnings distribution in different data sets play a role in the analyses and explain some differences in volatility levels and trends, as summarized below and analyzed in detail in the papers. All papers work with a sample of men who worked in both years, but results are also obtained when men with zero earnings in one year are included and which therefore capture volatility in the movement into and out of employment. Results including nonworkers are summarized in Appendix B.

IV. Results

A. Baseline

Figure 1 shows our baseline results, using the samples and earnings variables listed in Table 1, for men working both periods.[14] The PSID shows patterns mostly consistent with prior work, with rising volatility from the 1970s to the mid-1980s, then following a stable trend around significant fluctuations through about 2002, then rising in the period leading up to and including the Great Recession, and then falling post-Recession from 2012 to 2016.[15] The last four years of PSID data are new to this project and show that volatility has declined back to its pre-Recession level in 2006, which was somewhat above its level in the mid-1980s.

---

[13] All decompositions into permanent and transitory components require identifying restrictions on what goes into each component and, indeed, the general linear ARMA model is not identified (Moffitt and Zhang, 2018, Online Appendix). However, as noted in the Conclusions, future work on comparisons of trends in permanent and transitory variances in different data sets would be of interest

[14] The four individual papers are Carr et al. (2021), McKinney and Abowd (2022), Moffitt and Zhang (2021), and Ziliak et al. (2021).

The series for the other five data sets are often different from the PSID in level but not always in trend. The SIPP administrative data series, which starts in 1980, is higher in level than the PSID but follows a similar slight decline from 1982 through about 1999, but with fluctuations over that period much milder in magnitude than for the PSID (perhaps the result of a larger sample size—see Table 1). It then also rises before the Great Recession and falls afterwards, although again not always of the same magnitude and at exactly the same time points as the PSID. The SIPP survey data have a lower level of volatility than the PSID (and much lower than the SIPP administrative data series) but has an approximately similar pattern in the first half of the period—a rise then fall from 1985 to 1999, but occasionally moving in opposite directions (e.g., 1988-1990).[16] But the main difference with the SIPP survey is that it rises much less before and during the Great Recession than the PSID and the SIPP administrative data series.[17]

The two CPS series shown—one for survey data and one for administrative data—are computed only on non-imputed observations (see the CPS paper). The series are lower in level than the PSID but their separate levels are quite close to one another, an important finding suggesting that any response error in the CPS survey is small enough to be ignored, at least for the purpose of earnings volatility measurement. Both CPS series only begin in 1996 and, over that period, both rise with the Recession and then fall afterwards, returning to their original 1996 levels by 2015, implying no net trend. This differs from the PSID, which is still somewhat higher by that date than it was in 1996. Finally, the LEHD, which also only starts in 1998, has the highest volatility level of all the series. In terms of trends, it has two countercyclical spikes which leaves its value in 2011 the same as its initial value in 1998, but declines from 2011 to 2016 and ends up below its 1998 level. Most other series also declined after 2011 but not to values below their 1998 levels.[18]

---

[15] The PSID fluctuations are somewhat countercyclical, but there are many which do not line up well with unemployment rates and may be the result of sampling error. There is a special hypothesis for the large fluctuations in the 1990s having to do with data issues which we discuss below in footnote 19.
[16] The SIPP survey calculations use only non-imputed observations. See the SIPP paper.
[17] Several years of the SIPP survey are missing because of its irregular cohort design (see Table 1), and intentionally left blank in the Figure because there could be spikes in those years.

Overall, the differences in volatility levels across the series are greater than their differences in trends. As we will describe below, we are able to greatly narrow the differences in levels across the data sets. As for trends, the series are consistent with each other in many major respects. Prior to 1998, there is rough consistency across the three data sets we have for that period—essentially, no average trend in volatility from the mid-1980s to the 1998-2002 period. After 1998, the two CPS series, the SIPP administrative data series, and the PSID all show increases prior to and including the Great Recession, followed by declines. The magnitude of the upward trend is greatest for the PSID. However, the SIPP survey and the LEHD show little or no trend, and we will have some hypotheses for the differences with these two data sets below.

We devote a brief additional discussion to the post-1998 trends to explore these differences a bit further. There is always an issue in comparing growth rates of any short aggregate time series that has significant fluctuations without a formal statistical model because calculations of those rates can be highly sensitive to the chosen starting point. For example, earnings volatility growth in the PSID is much faster than in other data sets from 1998, but 1998 was a low point that clearly deviated negatively from its trend.[19] Both SIPP series have a dip in 1998 and the two CPS series have a dip in 2000 which suggests not using them as a starting points. While not claiming any particular formal justification for our procedure, we address this issue by measuring volatility growth after 2002 relative to the 1992-2002 average for the PSID and the SIPP administrative data, and relative to the 1998-2002 average for the other data sets. These particular intervals roughly average over one complete cycle for each data set, as can be seen from Figure 1. The results, appearing in Appendix Figure 1, show, first, that four of the data sets have remarkably similar growth and decline patterns after 2002 relative to their initial averages, and end up with a net zero growth by the end. A fifth, the PSID, is an outlier and is mainly distinguished by a continued high level after 2012 and a lack of decline in that post-Great-Recession period compared to the pattern exhibited by the other data sets, resulting in a net positive growth by the end (but the PSID is not much different than the others prior to 2012). The LEHD is the other outlier, showing very little cyclical growth and a stronger decline post-Recession than in the other data sets and ending with a small negative growth after about 2011. We conclude that average volatility growth rates in the 2000s and partway into the 2010s were quite similar for most of the data sets, with some exceptions. Combined with our finding of very little volatility trend in the three data sets covering the period from the mid-1980s to the 1998-2002 period, we also conclude that there is little evidence for significant trends upward or downward over the last 20-30 years.

---

[18] Most of the series' spikes in the Great Recession are at approximately the same year except for the PSID, whose spike occurs later. This may be partially a result of the two-year interval used in the PSID instead of the one-year interval used in the other data sets, although shifting the PSID spike back by one year would still leave it occurring later than those in the other data sets.

---

[19] As discussed in the PSID paper, there was a change in data collection and imputation procedure in the PSID over this period which we and Dynan et al. (2012) believe may be responsible to the extreme fluctuations in the 1990s.

B. <u>Explaining the Differences</u>

The difference across the six data series which we find to be most important in explaining their level and trend differences is related to differences in their cross-distributional distributions of earnings.   Before we present those results, we briefly summarize the large number of other hypotheses we have explored which have little or no explanatory power for differences in volatility trends.   The details of these investigations are in the individual papers; here we just summarize the findings.

For example, the restriction of the PSID to household heads is explored by estimating volatility trends in the CPS and SIPP survey, whose results above include non-heads, on heads only.   Both data sets show lower levels of volatility for heads than for non-heads but trends are unaffected (Carr et al., 2021, Appendix Figure A.5; Ziliak et al., 2021, Appendix Figure S.9). For the differences in volatility trends for wage and salary earnings, both the SIPP and CPS permit the estimation of volatility excluding the self-employed, as required in the PSID.   The SIPP data show a lower level of volatility for wage and salary earners but no change in trend, while the CPS has similar findings (Carr et al., 2021, Appendix Figure A.6; Ziliak et al., 2021, Appendix Figure S.9).   For the PSID exclusion of immigrants, the CPS paper finds no volatility trend differences for immigrants and non-immigrants (Ziliak et al., 2021, Appendix Figure S.8). And regarding our baseline regression specification, volatility trends are little affected by using regression residuals from log earnings equations instead of log earnings itself, or by using the log earnings difference instead of the arc percent change (Moffitt and Zhang, 2021, Appendix Figure 2; Carr et al., 2021, Appendix Figures A.1 and A.2; Ziliak et al., 2021, Appendix Figure S.3 and Figure S.6; McKinney and Abowd, 2022, Figures 1 and Appendix Figure B2).[20]   In addition, our trimming at the 1st and 99th percentile points of the cross-sectional earnings distributions has no effect on trends; no trimming at all produces more fluctuations in our estimated trends and trimming greater proportions of the tails produces fewer fluctuations, but in neither case are trends affected (Moffitt and Zhang, 2021, Appendix Figure 7; Carr et al., 2021, Appendix Figure A.3 and A.4); Ziliak et al., 2021, Appendix Figures S.4 and S.5; McKinney and Abowd, 2022, Figure 1).

---

[20] In addition, we find that, while the trends are the same for the measures, the log earnings variance has more fluctuations around the trend.   This is to be expected since the log measure puts heavier weight on values in the tail.

The papers in the project note that it is important to trim at percentile points and not to use real dollar trims, as employed in some prior work using administrative data (Kopczuk et al. (2010), Guvenen et al. (2014), Bloom et al. (2017); Sabelhaus and Song (2009, 2010)). As noted by Carr and Wiemers (2021), using real dollar trims creates bias if either the tails of the distribution are changing in real dollar terms or if the trends in volatility are different in the tails. Appendix A to this Overview reports the results of our analysis of this issue with our different data series and shows that, in some data sets, real dollar trims sometimes reduce the upward trends in volatility and, in some cases, change an upward trend to a negative trend.

Attrition is more of a potential threat because it only occurs in survey data sets.[21] Bias in trends can occur if those who are missing from a survey have differences in volatility, and attrition in our survey data sets range from 20 to 44 percent. For one of our data sets—the CPS—we have administrative data for those who attrite. CPS attrition takes place when a family interviewed in the first year has moved or is otherwise unavailable in the second year. The CPS analysis shows that those missing in the second year have much higher levels of volatility than those who are not missing but that trends are unaffected when they are included (Ziliak et al., 2021, Figure 4). In addition, in all three survey data sets--the CPS, SIPP, and PSID—we use standard inverse probability weighting to test for attrition bias by estimating the probability of attrition as a function of observables and then reweighting the volatility calculation on the non-attriter sample with the inverse of the predicted probabilities (Wooldridge, 2010). This eliminates bias under the selection-on-observables assumption. The results show virtually unchanged trends in volatility after this adjustment (Moffitt and Zhang, 2021, Appendix Figure 10; Carr et al., 2010, Appendix Figure A8; Ziliak et al., 2021, Figure 5). The results for the PSID, in fact, show a stronger upward trend in volatility after reweighting, consistent with prior studies of the PSID indicating that higher-volatility individuals are more likely to drop out in the first place (Fitzgerald et al., 1998).

---

[21] While it is possible for individuals with different levels of volatility to drop out of administrative data sets (e.g., by moving to unreported earnings jobs), the magnitude of the problem is much smaller than in survey data sets.

In addition to attrition, a fraction of respondents in all surveys have missing data on specific variables ("item nonresponse") because of do-not-know responses and refusals-to-answer, from implausible values indicating response error, and for other reasons. Surveys typically impute new values for those missing responses and the statistical properties of those imputations have attracted a great deal of discussion in the literature (e.g., Andridge and Little, 2010). While imputation rates for earnings in the PSID are very low—3 to 4 percent—they are very high in the CPS and SIPP, with recent rates of about 45 and 49 percent of all observations, respectively, and rising over time. Further, for both the CPS and SIPP, studies linking survey data to administrative earnings reports show that missingness is decidedly non-ignorable, with nonresponse higher in the tails of the earnings distribution (Bollinger et al., 2019; Chenevert et al., 2015). The non-ignorability is important because imputation methods for the CPS and SIPP made by the U.S. Census Bureau only use observables to impute values for the missing observations. For our project, while more sophisticated approaches to the problem are possible, we take simple approaches for the CPS and the SIPP. The CPS analysis shows that volatility levels and trends are very different when using the administrative earnings data for those whose earnings are imputed. Specifically, using imputed survey earnings show much higher levels of volatility as well as a rising trend (Ziliak et al., 2021, Figure 3). But, as noted previously, both the level and trends in volatility are virtually identical in survey and administrative data when using only non-imputed survey earnings.

The SIPP analysis cannot link administrative data directly to the survey data but it is apparent that imputation is a serious problem in the survey data. Methods of imputation and their coding have changed over time, which makes a truly consistent measure of imputation not possible with the SIPP survey. Carr et al. (2021, Figure 2) shows that the more serious problem is the presence of so-called "whole imputes," where the entire observation is imputed. Including those observations (which were excluded for the calculations shown in Figure 1) greatly raises the level of volatility as well as changing the flat trend to a positive one.

<u>Differences in Cross-sectional Earnings Distributions</u>. It has been noted in a number of prior studies that administrative data on earnings from Social Security and UI earnings records appear to have larger left tails of the earnings distribution than survey data sets (Kornfeld and Bloom, 1999; Juhn and McCue, 2010; Celik et al., 2012; Spletzer, 2014; Abraham et al., 2013; Abowd et al., 2018). The most common hypothesis for this difference is that survey respondents often fail to report earnings from short, part-year jobs, especially if asked about earnings over a previous 12-month period.[22] This difference is strongly exhibited in our six data series, as shown by the cumulative distribution functions in Figure 2, which shows the LEHD to have the largest left tail of earnings.   The left tail of the SIPP administrative data is larger than that of the SIPP survey data and the left tail of the CPS administrative data is larger than that of the CPS survey data.   The PSID has the smallest left tail.   For example, about 25 percent of the LEHD observations have earnings less than $20,000 per year while only about 5 percent of PSID observations do, a large difference.   These differences will almost surely cause the levels of volatility to be larger in administrative data because the literature has shown volatility to be higher at lower earnings levels than at higher ones.   Figure 1 shows volatility levels indeed to be higher in administrative data than in survey data.   But it can affect trends in volatility as well as levels if either those trends differ between the lower tail and the rest of the earnings distribution or if the size of the left tail is changing differently in the different data sets.

---

[22] However, matches of other surveys to administrative data have found differences in both directions—survey reports typically have jobs and earnings reports that are missing from the administrative data as well as the other way around. In many cases, this seems to be because the administrative data are in error and do not, for a variety of reasons, pick up jobs and earnings that survey respondents report (Juhn and McCue, 2010; Abraham et al., 2013; Abowd and Stinson, 2013).

We explore this issue by adjusting all data series to the same cross-sectional distribution. Given the key role of the PSID in this literature, we benchmark the other five to it, which effectively means downweighting the left tails of the other five data sets' distributions.   We do so by first assembling the minimum and maximum value of the PSID cross-sectional earnings distribution in each year (which are the $1^{st}$ and $99^{th}$ percentile points of the untrimmed distribution) and then computing ventile percentile points in each year between those year-by-year minima and maxima.   For each of the other five data series, we discard observations below the PSID minimum and above the PSID maximum in each year, and then use the PSID ventile points to compute a weighted average volatility, weighting the observations in each ventile range by .05. The results will reveal whether any differences in volatility levels and trends across the data series arise from differences in their cross-sectional earnings distributions rather than differences in volatility trends conditional on location in the cross-sectional distribution (at least within the PSID range).

Figure 3 shows the benchmarked volatility for the five series other than the PSID, as well as the PSID for comparison. The benchmarking has a dramatic impact on the comparisons between data series, as can be seen by a comparison to Figure 1. The levels of volatility for the three administrative data sets (CPS administrative, SIPP administrative, and LEHD) are greatly lowered because of the trimming and/or downweighting of their large left tails. Now the levels of all data sets except the PSID are relatively tightly concentrated in a narrow range in the 1998-2015 period (they are all still quite a bit below the PSID level). The trends for the five also show similar time patterns in their overlapping years, rising from the mid-1990s through the peak of the Great Recession and declining thereafter, as in Figure 1, except that now the LEHD trend is a small net positive from the first year to the last instead of a small net negative. The changed trend in LEHD volatility arises because the PSID real $1^{st}$ percentile point declines over time while the lower percentile points in the LEHD do not, leading to the inclusion of an increasing fraction of (high volatility) low LEHD earners over time (McKinney and Abowd, 2022, Figure 3 and Appendix Figure G1). Thus the underlying source of the change is a difference in trends in the left tails of the earnings distributions.

Finally, to compare trends after the 1990s, we can benchmark all six data sets to their initial average as discussed above but using the reweighted data series instead of the unweighted series. As shown in Appendix Figure 2, the LEHD series ends with a net small positive growth in volatility but which is close to zero, and the CPS survey follows an almost identical pattern. The SIPP survey follows a pattern similar to these two, over the years it is available. The CPS and SIPP administrative series do not fall as much—because of the downweighting of their left tails—and now follow trends almost identical to that of the PSID. We conclude that the majority of the large differences in volatility levels across the data series, and much of the small differences in trend, are explained by the differences in their cross-sectional earnings distributions.

V. Summary and Conclusions

The project which this Overview summarizes has been narrowly focused on the sole question of how the gross earnings volatility of prime-age men has evolved over the last 40-50 years in the US and whether differing findings on this question across different data sets can be reconciled. The central finding is that, when put on a comparable basis, male earnings volatility in six survey and administrative data sets shows no sign of major net increase or decrease since the late 1980s or early 1990s, although experiencing significant countercyclicality. There is some evidence that volatility increased from the 1970s to the 1980s but only from a subset of our data series which go back that far.   Our findings should be regarded as a significant contribution to our understanding of the evolution of the US labor market.

The narrowness of the exercise means that many interesting questions have not been addressed. Extensions to other demographic groups, by age and gender and marital status, would be warranted. There is no reason that labor market volatility trends for those groups should be the same as for prime-age men. Decomposing male earnings volatility into its components of hours worked and wage rates and job mobility would also be of interest.   Error components models exploring the dynamic structure of earnings evolution could yield additional insights on time-series trends in all those components. And an important finding of our work is that earnings volatility is quite different in different parts of the earnings distribution, which suggests that future work address that source of heterogeneity in whatever outcomes are studied.

But a lesson from our work is that all data sets have their strengths and weaknesses, and acknowledging those characteristics and investigating how they may affect the results of a particular study should be an important goal for work in this area.

Appendix A

Real Dollar Trims

We follow the suggestion of Carr and Wiemers (2021) that a number of papers in the volatility literature using Social Security earnings records have trimmed the earnings distribution at the bottom of the distribution using real-valued cutoffs instead of percentile points, and that this could affect measured trends in volatility in some cases. For example, if the earnings distribution is symmetric around a constant but dispersion and inequality are growing over time, a constant real-dollar trim will trim out an increasing fraction of the tails, affecting average volatility if volatility differs across the cross-sectional earnings distribution (typically, volatility is higher at lower earnings levels). Carr and Wiemers point out that some studies have also trimmed at the bottom of the distribution on real dollar values that change over time, which could have further complicated effects on average volatility.

Following Carr and Wiemers, we test the effect of excluding observations with real annual earnings below one quarter of full-time full-year work at the 2011 federal minimum wage, a method used by Kopczuk et al. (2010) (this is $3,685 in our 2010 dollars); excluding observations with real annual earnings below a quarter of a year of full-time full-year work at half the federal minimum wage, but using the actual minimum in each year, a method used by Guvenen et al. (2014) and Bloom et al. (2017); and excluding observations below the annual earnings needed to qualify for the Social Security threshold for credit, a method used by Sabelhaus and Song (2009, 2010).

Applying these trims to the PSID results in a slower growth in volatility over the entire period, from 1972 to 2018, for the first and third methods (Moffitt and Zhang, 2021, Figure 2). The profiles are especially flatter in the 2000s and now show no growth in volatility at all after about 2000. For the SIPP administrative data, somewhat greater rates of decline or slower rates of increase than before result when using the first and third methods (Carr et al., 2021, Appendix Figure A.3). But, for the CPS, SIPP, and LEHD, the three trimming methods affect the levels of volatility but not the trends (Ziliak et al., 2021, Figure S.10; Carr et al., 2021, Appendix Figure A.4; McKinney and Abowd, 2022, Appendix Figure D2). The reason for the difference in effects across the data sets appears to arise from differences in whether real earnings inequality is rising or stationary. When it is rising, as it is for the PSID and the SIPP administrative data, the real dollar trimming makes a significant difference, generally generating a slower upward trend or faster downward trend. This is not surprising if the real dollar lower tail is growing over time. But real wage inequality is increasing very little in the other data sets, which leads to little effect of real dollar trimming on volatility trends.

Appendix B

Including Nonworkers

All the central analyses have utilized a sample of men working at both points over time for volatility measurement. Nonworkers are therefore excluded. If nonworkers are included and simply treated as zero earners, and earnings changes are calculated as the difference in earnings between the two periods, including nonworkers will probably increase the level of volatility.[23] However, there is no presumption of how including nonworkers would affect volatility trends, because that depends on whether there is a trend in the transition rates into and out of work. For example, a reduction in the cross-sectional employment rate over time (which has, indeed, been occurring for men) has no implication for trends in volatility; volatility could even decline if, for example, those who have moved into nonwork stay in that status without ever, or rarely, transitioning out of it. Average volatility over worker and nonworker earnings changes could consequently fall relative to including workers only in this case.

When including nonworkers, most studies in the literature use the arc percent change (APC) measure of volatility rather than log earnings because the latter does not allow zeros, at least not without modification of the log earnings function. The APC does allow the inclusion of zeroes, at least if earnings are zero for only one period (those with zero earnings both periods still must be excluded).[24] But while including nonworkers is straightforward in surveys, it is less so in administrative data, which consist only of earnings records. To include zero earners in volatility calculations using administrative data, the absence of an earnings record is treated as a zero earnings observation. But there may be other reasons for the absence of a wage record, such as working off the books, working at an uncovered firm, and for other reasons.

---

[23] We say "probably" because there is no mathematical requirement for this to be the case, because the magnitudes of year to year changes in earnings could be larger than those of earnings changes to and from 0 (e.g., if movements in and out of work are concentrated among those with low earnings levels). However, this is unlikely to hold, and it does not hold in our data sets.

[24] A more formal econometric approach would be to specify a two-equation model with a binary choice equation for employment and a second equation for latent earnings which is only realized if employment occurs. We leave such a model for future work.

For our project, APC volatility results have been calculated for the PSID, CPS, and SIPP (Moffitt and Zhang, 2021, Appendix Figure 2; Ziliak et al., 2021, Figure 2; Carr et al., 2021, Appendix Figure A7). In all cases, the level of volatility is higher when nonworkers are included and, in all cases, the countercyclicality of volatility is much greater. More important, including nonworkers makes volatility growth more positive on net—that is, ignoring business cycle movements--at least after the early 2000s (including nonworkers has little effect on trends before then). Further, while volatility including nonworkers declines after the Great Recession for all data series, for the CPS and SIPP (both survey and administrative) it ends up at a higher level than in the early 2000s—as opposed to the baseline results which showed them to end at the same level--while, for the PSID, it declines much more than in the baseline results and ends up at a level below all years back to 1990.

…

Table 1: Comparisons of the 6 Data Series

|  | Years | Sampling Frame | Sample | Earnings Variable[1] | Nonworkers | Approximate Sample Size[2] |
|---|---|---|---|---|---|---|
| PSID | 1970-2018 | Representative Sample of Families in 1968 and their descendants | Male heads of household or spouses of head, 25-59 | Wage and salary earnings (self-employment not available) | Available | 1,027 |
| SIPP-Survey | 1984-2012[3] | Representative Sample in initial year of each panel | Men 25-59 including non-heads (headship measurable) | Wage and salary and self-employment separately | Available | 4,358 |
| SIPP-GSF | 1980-2014 | All individuals in SIPP families in 1984 and 1990-2008 panels who have valid SSN | Men 25-59 (headship not measurable) | Total earnings from all jobs with a W-2 or a Schedule C (self-employment) (cannot separate self-employment) | Mostly available | 103,667 |
| CPS-ASEC | 1995-2015 | Individuals matchable across 2 annual CPS surveys | Men 25-59 including non-heads (headship measurable) | Wage and salary and self-employment separately | Mostly available | 5,400 |
| CPS-DER | 1995-2015 | Those in the ASEC sample who can be found in SS earnings records | Same as ASEC | Same as ASEC, though self-employment restricted to positive earnings | Mostly available | 5,600 |
| LEHD | 1998-2016 | UI earnings records for jobs associated with a valid SSN[4] | Men 25-59, including non-heads (headship not measurable) | Wage and salary, self-employment only if reported in UI | Not available | 47 million |

[1] Tips, bonuses, and commissions are included in all data sets except the PSID, although tips are probably underreported in administrative data.
[1] Average number of paired positive earnings observations per year pair.   [3] Years 1990, 1996, 2000, 2001, 2008 and 2009 missing. [4]Selected states.

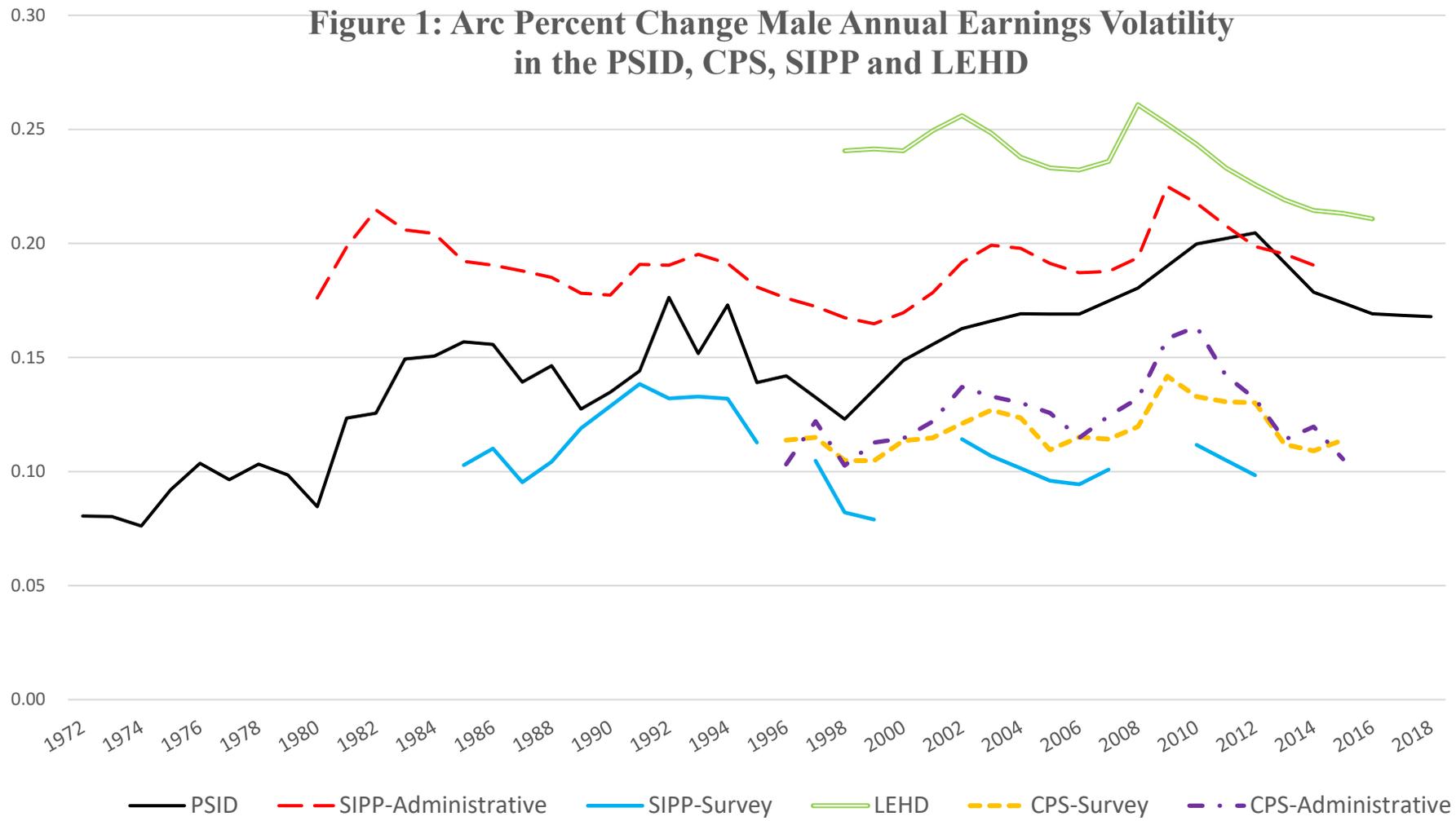

*Notes*: Sample consists of men with positive earnings in both time periods with a 1 percent top and bottom trim of the cross-sectional earnings distribution in each year. Moffitt and Zhang (2021, Appendix Figure 2). CPS: Ziliak et al. (2021, Appendix Figure S.6). SIPP: Carr et al. (2021, Appendix Figure A2). LEHD: McKinney and Abowd (2022, Figure 1).

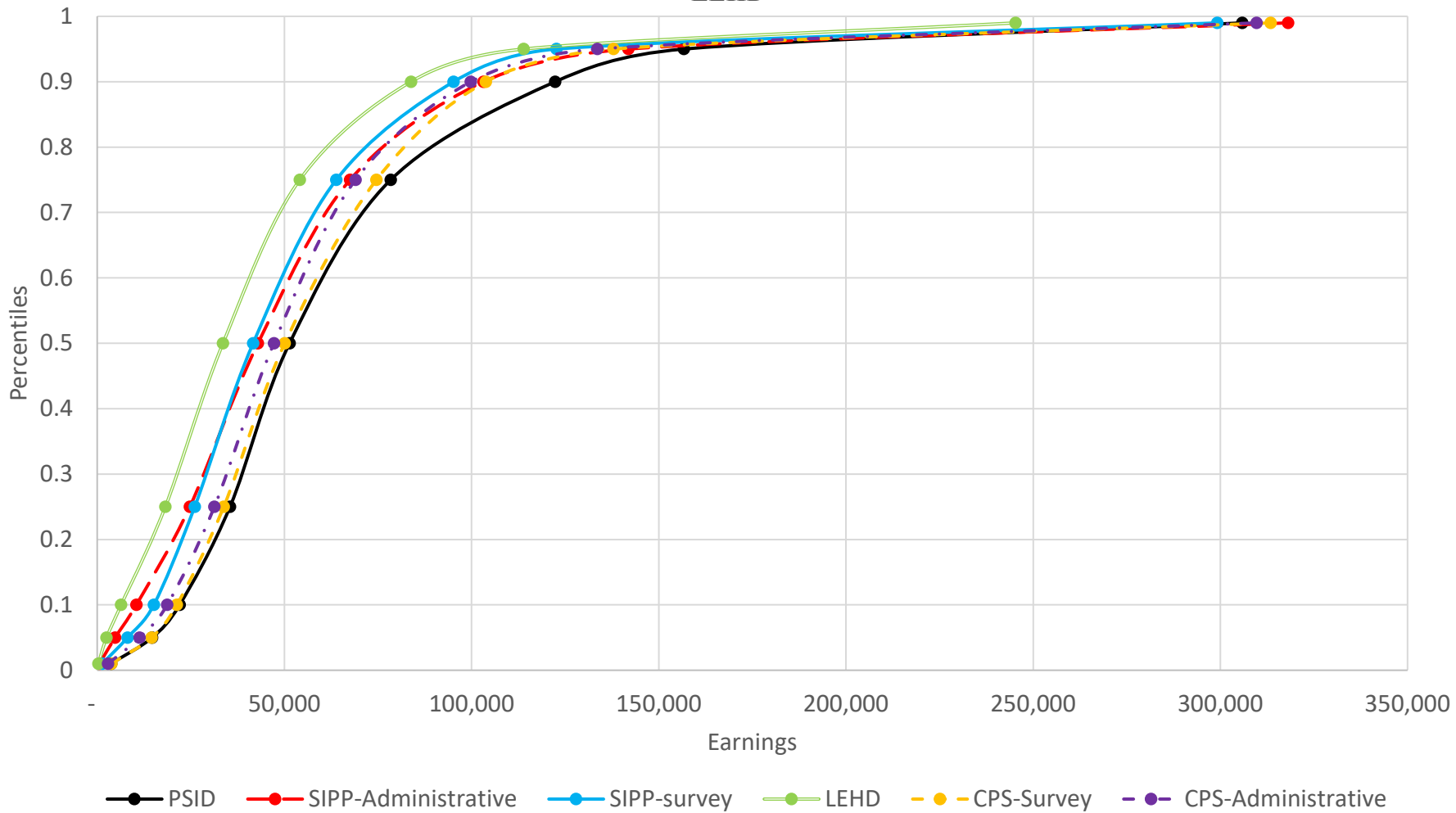

*Notes*: The PSID, CPS-Survey, CPS-Administrative use 2000 data, and SIPP-Administrative, SIPP-Survey, and LEHD use 2001 data.

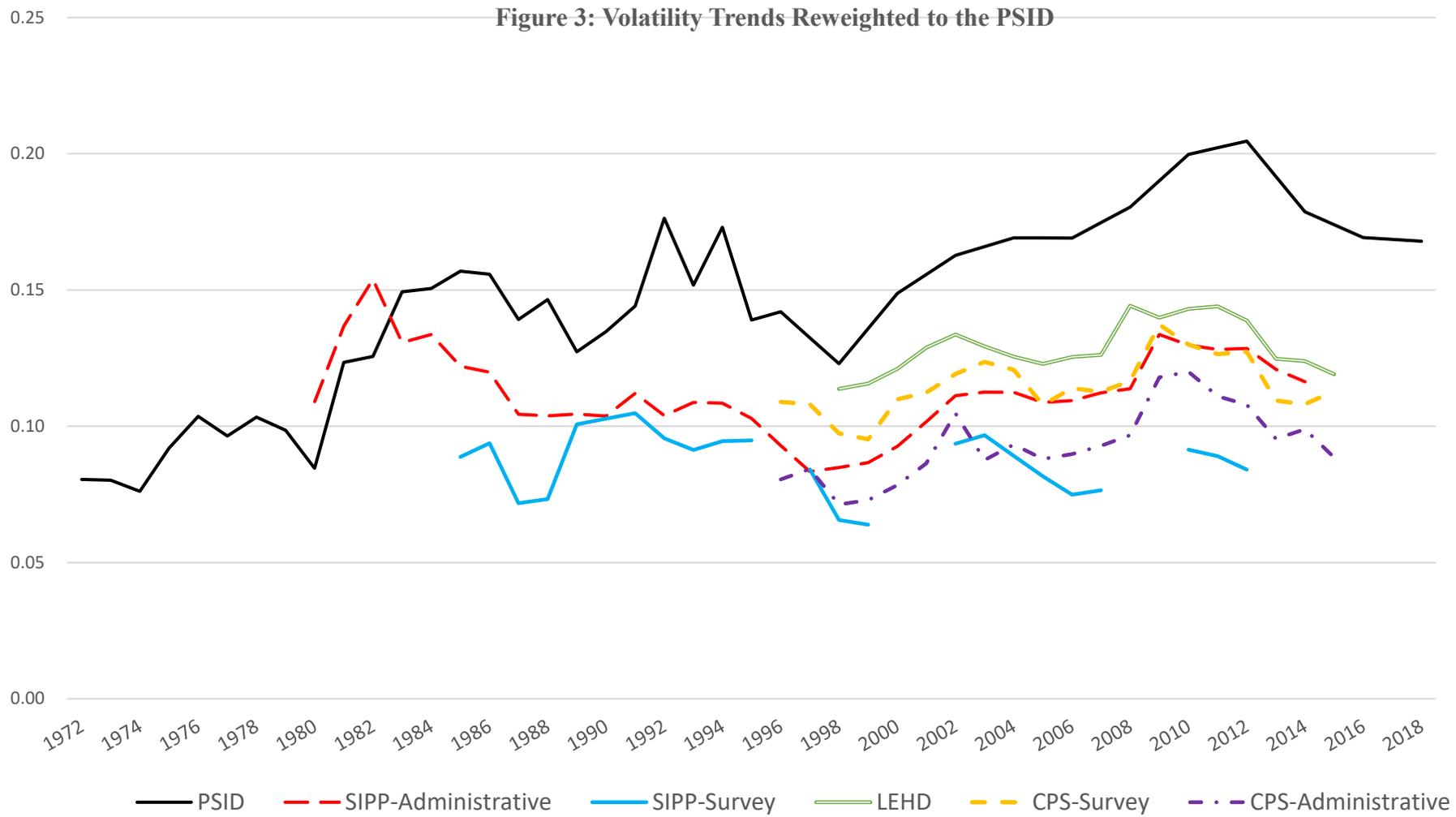

*Notes*: Sample consists of men with positive earnings in both time periods with a 1 percent top and bottom trim of the cross-sectional earnings distribution in each year. Cross-sectional distribution of all data sets except the PSID reweighted to match the PSID cross-sectional distribution. CPS: Ziliak et al. (2021, Appendix Figure S.11). SIPP: Carr et al. (2021, Figure 3). LEHD: McKinney and Abowd (2022, Figure 3).

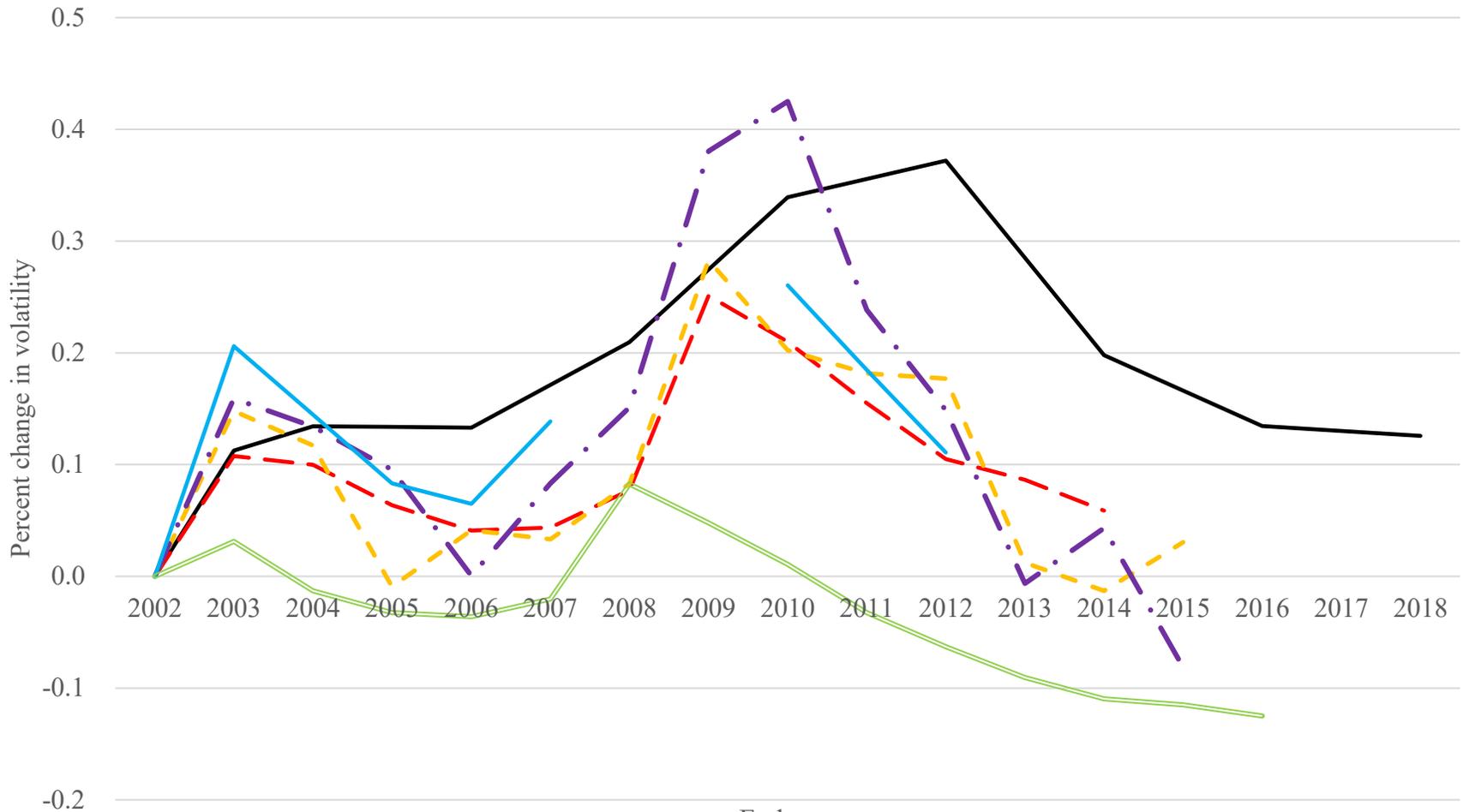
Appendix Figure 1: Post-2002 Percent Change in Volatility Relative to Initial Average

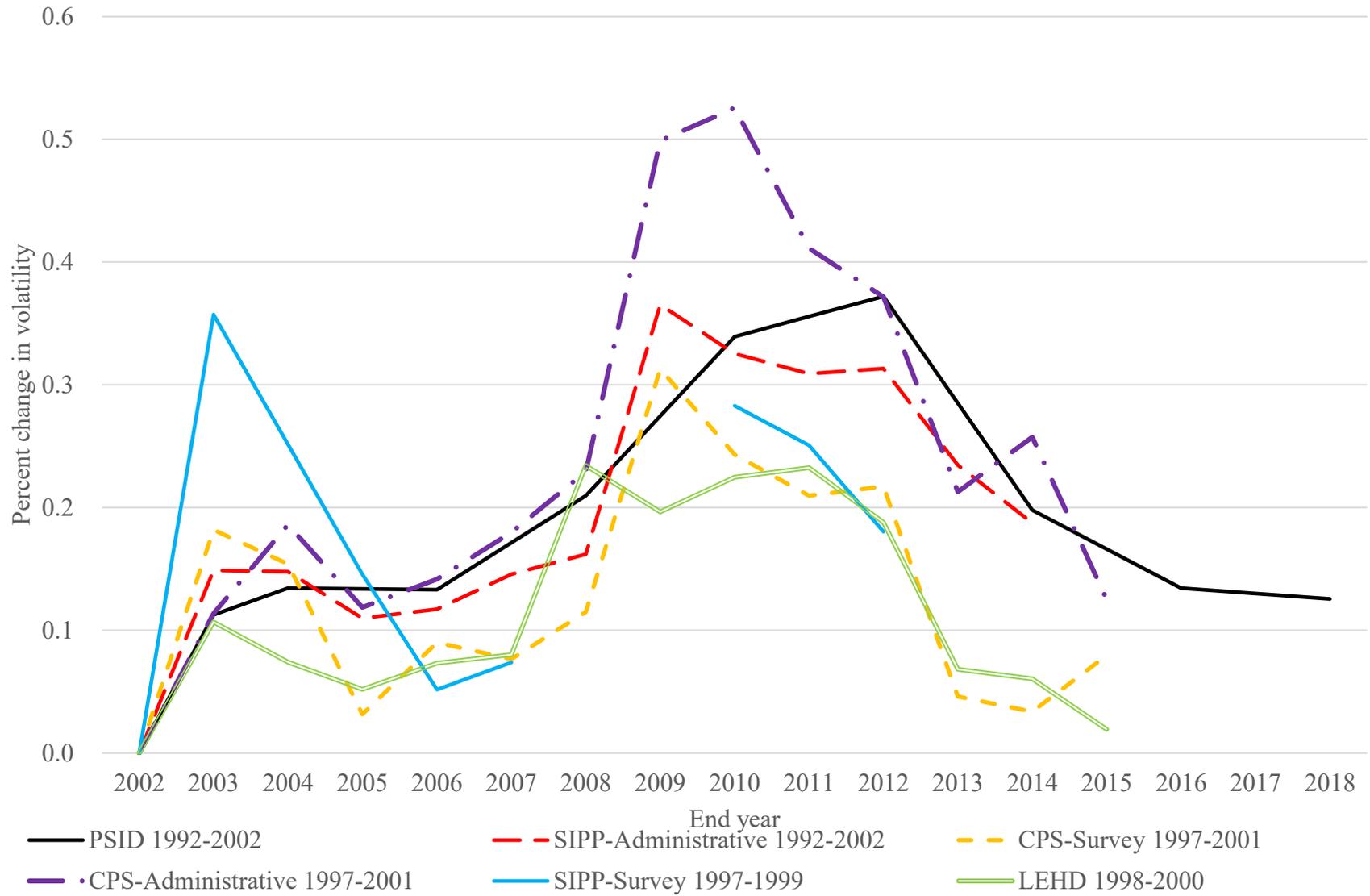

Appendix Figure 2: Post-2002 Percent Change in Volatility Relative to Initial Average, Data Reweighted to PSID